\documentclass[10pt,a4paper]{article}

\usepackage[utf8]{inputenc}
\usepackage[english]{babel}

\begin{document}
\large

\newpage
\begin{center}
\LARGE{\bf On the Structure of Leptonic Families and Currents 
\\of the Vector Nature}
\end{center}
\vspace{0.1cm}
\begin{center}
{\bf Rasulkhozha S. Sharafiddinov}
\end{center}
\vspace{0.1cm}
\begin{center}
{\bf Institute of Nuclear Physics, Uzbekistan Academy of Sciences,
\\Tashkent, 100214 Ulugbek, Uzbekistan}
\end{center}
\vspace{0.1cm}

\begin{center}
{\bf Abstract}
\end{center}

Any of neutrinos similarly to a kind of charged lepton has a non-zero mass responsible as 
well as for its Coulomb behavior. Such a neutrino can possess both electric charge and vector 
dipole moment. Their form factors appear, for example, at the polarized neutrino scattering 
in the field of a spinless nucleus. We derive an equation which relates the masses to a ratio 
of Dirac and Pauli form factors of each lepton and its neutrino. A new left-right antisymmetric theory of force unification based on a gauge group 
$SU(2)_{L}$ $\otimes$ $SU(2)_{R}$ $\otimes$ $U(1)$ is suggested. In this theory, the leptons
and their neutrinos are united in families not only of the left-handed $SU(2)_{L}$-doublets, 
but also of the right-handed $SU(2)_{R}$-doublets. Thereby, it predicts the existence in nature 
of the left (right) dileptons and paradileptons. A formation of any of them is responsible for 
the legality of conservation of charge, lepton flavors and full lepton number. Therefore, each 
of the earlier measured processes originated at the conservation both of summed electric charge 
and of any lepton number may serve as the first confirmation of a given theory in which the mass, 
charge and vector dipole moment of the neutrino are proportional, respectively, to the mass, 
charge and vector dipole moment of a lepton of the same family.

\vspace{0.6cm}
\noindent
{\bf 1. Introduction}
\vspace{0.2cm}

In studying the structure of fundamental interactions of electrons and their neutrinos with 
the field of emission such characteristics as the Dirac $F_{1l}(q^{2})$ and Pauli $F_{2l}(q^{2})$ 
form factors [1] play a large role. An explicit form of these functions thus far remains not 
finally established. It is usually accepted that form factors $F_{1l}(q^{2})$ and $F_{2l}(q^{2})$ depending on the momentum transfer square $q^{2}$ correspond to the electric and magnetic 
components of the vector leptonic current. However, according to the micro-world symmetry 
laws, the presence of any type of an electrically charged particle implies the existence 
of a kind of magnetically charged monoparticle [2]. It is fully possible therefore that the corresponding mononeutrinos lead to quantization of the electric charges of all neutrinos 
and vice versa. Under such circumstances, well known electromagnetic field is described as 
the field of the unified system of the photon and monophoton where the two forces of the 
electric and magnetic nature are united [2].

Furthermore, if it turns out that a non-zero interaction of Pauli arises as a consequence of the availability of Dirac interaction [3], the function $F_{2l}(q^{2})$ cannot be magnetic form factor. To characterize it from the point of view of an electrically charged particle $(l=e,$ $\mu,$ 
$\tau, ...$ or $\nu_{l}=\nu_{e},$ $\nu_{\mu},$ $\nu_{\tau}, ...)$ one must use a new tongue. We 
need therefore to call $F_{2l}(q^{2})$ an electric vector moment. Similar terminology becomes  unsubstantiated for a magnetically charged monolepton $(l=l_{H})$ when $F_{2l_{H}}(q^{2})$ 
fulfils the function of magnetic form factor of this monofermion.

The difference in nature of elementary particles and monoparticles appears
because of the duality that the mass and charge correspond to the two form
of the same regularity of the structure of matter [4].

Such a sight to the appearance of an intimate connection between the mass of the neutrino and 
its electric charge [5] quality explains the fact that both mass $m_{l}$ and charge $e_{l}$ of 
a particle are strictly multicomponent. Any of them contains as well as the electric $(E),$ weak $(W)$ and strong $(S)$ parts. They constitute herewith the naturally united rest mass $m_{l}^{U}$ and charge $e_{l}^{U}$ of the same particle [4] equal to all its mass and charge
\begin{equation}
m_{l}=m_{l}^{U}=m_{l}^{E}+m_{l}^{W}+m_{l}^{S}+...,
\label{1}
\end{equation}
 \begin{equation}
e_{l}=e_{l}^{U}=e_{l}^{E}+e_{l}^{W}+e_{l}^{S}+....
\label{2}
\end{equation}

This picture reflects the availability of an interratio of intralepton forces of a different 
nature. Therefore, each of form factors $F_{il}(q^{2})$ comes forward in the system as the 
Fourier transform of the spatial density of the corresponding interaction. One can present [3]
their in the form 
\begin{equation}
F_{il}(q^{2})=f_{il}(0)+A_{il}(\vec{q^{2}})+...,
\label{3}
\end{equation}
where terms $f_{il}(0)$ define the normal values of the electric charge and vector dipole moment:
\begin{equation}
f_{1l}(0)=e_{l}^{norm}, \, \, \, \, f_{2l}(0)=\mu_{l}^{norm}.
\label{4}
\end{equation}
Here the size of $e_{l}^{norm}$ for the lepton (antilepton) has the negative (positive) sign.

The functions $A_{il}(\vec{q^{2}})$ characterize the anomalous components of form factors, the values of which for $\vec{q^{2}}=0$ must be equal to 
\begin{equation}
A_{1l}(0)=e_{l}^{anom}, \, \, \, \, A_{2l}(0)=\mu_{l}^{anom}.
\label{5}
\end{equation}

At these data, $F_{1l}(0)$ and $F_{2l}(0)$ coincide with the full static electric charge 
and vector dipole moment:
\begin{equation}
F_{1l}(0)=e_{l}=e_{l}^{full}=e_{l}^{norm}+e_{l}^{anom}+...,
\label{6}
\end{equation}
\begin{equation}
F_{2l}(0)=\mu_{l}=\mu_{l}^{full}=\mu_{l}^{norm}+\mu_{l}^{anom}+....
\label{7}
\end{equation}

One of the fundamental regularities in nature of form factors is an individual connection 
between the corresponding characteristics of their structure [3]. They have the form
\begin{equation}
f_{2l}(0)=\frac{f_{1l}(0)}{2m_{l}^{norm}},
\label{8}
\end{equation}
\begin{equation}
A_{2l}(0)=\frac{A_{1l}(0)}{2m_{l}^{anom}},
\label{9}
\end{equation}
where $m_{l}^{norm}$ and $m_{l}^{anom}$ denote the normal and anomalous parts of all 
the electric mass
\begin{equation}
m_{l}^{E}=m_{l}^{norm}+m_{l}^{anom}+....
\label{10}
\end{equation}

According to the recent presentations about the nature of fermions, all leptons constitute 
the united families of the left-handed $SU(2)_{L}$-doublets and of the right-handed 
$SU(2)_{R}$-singlets:
\begin{equation}
\pmatrix{\nu_{e}\cr e^{-}}_{L},
e^{-}_{R}, \, \, \, \,
\pmatrix{\nu_{\mu}\cr \mu^{-}}_{L},
\mu^{-}_{R}, \, \, \, \,
\pmatrix{\nu_{\tau}\cr \tau^{-}}_{L},
\tau^{-}_{R}, ...,
\label{11}
\end{equation}
\begin{equation}
\pmatrix{{\bar \nu_{e}}\cr e^{+}}_{R},
e^{+}_{L}, \, \, \, \,
\pmatrix{{\bar \nu_{\mu}}\cr \mu^{+}}_{R},
\mu^{+}_{L}, \, \, \, \,
\pmatrix{{\bar \nu_{\tau}}\cr \tau^{+}}_{R},
\tau^{+}_{L}, ....
\label{12}
\end{equation}

However, a question about the leptonic families of $SU(2)_{R}$-doublets thus far remains 
open. It is usually assumed that unlike the $e_{R}^{-}(e_{L}^{+}),$ $\mu_{R}^{-}(\mu_{L}^{+})$
and $\tau_{R}^{-}(\tau_{L}^{+}),$ their right (left)-handed neutrinos have no neither weak, electromagnetic nor strong interaction, although possess a non-zero rest mass.

But in fact to each charged lepton corresponds in nature a kind of Dirac neutrino [6]. Their 
family must distinguish from others by the definite type of lepton numbers [7,8]. Herewith the 
three $(l=e,$ $\mu,$ $\tau.)$ numbers are rewritten to any particle:
\begin{equation}
L_{l}=\left\{ {\begin{array}{l}
{+1\quad \mbox{for}\quad l_{L}^{-}, \, \, \, \, \, l_{R}^{-}, \, \, \, \,
\nu_{lL}, \, \, \, \, \nu_{lR},}\\
{-1\quad \mbox{for}\quad l_{R}^{+}, \, \, \, \, l_{L}^{+},
\, \, \, \, {\bar \nu_{lR}}, \, \, \, \, {\bar \nu_{lL}},}\\
{\, \, \, \, \, 0\quad \mbox{for}\quad \mbox{remaining particles.}}\\
\end{array}}\right.
\label{13}
\end{equation}

Conservation of full lepton number
\begin{equation}
L_{e}+L_{\mu}+L_{\tau}=const
\label{14}
\end{equation}
or lepton flavors
\begin{equation}
L_{l}=const
\label{15}
\end{equation}
in the $\beta$-decay processes
\begin{equation}
n^{\pm}\rightarrow p^{\pm}e^{\mp}{\bar \nu_{e}}(\nu_{e}), \, \, \, \,
p^{\pm}\rightarrow n^{\pm}e^{\pm}\nu_{e}({\bar \nu_{e}}),
\label{16}
\end{equation}
\begin{equation}
\mu^{\mp}\rightarrow e^{\mp}{\bar \nu_{e}}(\nu_{e})
\nu_{\mu}({\bar \nu_{\mu}}), \, \, \, \,
\tau^{\mp}\rightarrow e^{\mp}{\bar \nu_{e}}(\nu_{e})
\nu_{\tau}({\bar \nu_{\tau}}),
\label{17}
\end{equation}
as well as in other reactions with neutrinos is by no means excluded experimentally [9]. But 
its legality has not yet been substantiated theoretically [10-12].

At our sight, such an order takes place owing to the unified nature, similarity and difference 
of each pair of the structural particles in families. Therefore, to understand the family structure 
of fermions at the fundamental dynamical level one must elucidate whether there exists a connection between the properties of leptons of the same families of doublets, and if so what an expected dependence says about the completeness of a vector picture of massive Dirac neutrinos and about 
the legality of conservation of charge, lepton flavors and full lepton number. The answer to this question one can obtain by studying the interaction with virtual photons of longitudinal light leptons. At such a situation, the processes on the nuclear targets [5,13,14] may serve as the 
source of unique information.

The purpose of a given work is to elucidate the nature of family structure of leptons 
investigating their Dirac and Pauli interactions with the field of emission in the presence of fermion longitudinal polarization. First of all a question about the appearance of an individual connection of the normal parts of the electric mass, charge and vector dipole moment of the 
neutrino and electron has been considered at the elastic scattering on a spinless nucleus. At 
the use of the interaction cross section with nuclei, the united dependence between the Dirac 
and Pauli form factors of the electron and its neutrino has been obtained. Next, we listed some consequences and logical implications implied from these discussions which give the possibility 
to directly look on the nature of conservation of charge and of each lepton number.

\vspace{0.6cm}
\noindent
{\bf 2. Individual Connection of Vector Currents of Neutrino and Electron}
\vspace{0.2cm}

The amplitude of elastic scattering of light leptons by nuclei in the limit
of one-photon exchange may be presented as
\begin{equation}
M^{E}_{fi}=\frac{4\pi\alpha}{q^{2}}\overline{u}(p',s')[\gamma_{\mu}
f_{1l}(0)-i\sigma_{\mu\lambda}q_{\lambda}f_{2l}(0)]u(p,s)
<f|J^{\gamma}_{\mu}(q)|i>.
\label{18}
\end{equation}
Here $\sigma_{\mu\lambda}=[\gamma_{\mu}, \gamma_{\lambda}]/2,$ $l=e=e_{L,R}$
or $\nu=\nu_{e}=\nu_{eL,R},$ $q=p-p',$ $p$ and $p'$ imply the four-momentum
of incoming and outgoing particles, $s$ and $s'$ denote their helicity,
$J_{\mu}^{\gamma}$ is the nucleus electric current [15].

For the case of spinless nuclei with an electric charge $Z,$ the cross section of the considered process on the basis of (\ref{18}) one can write in the form
$$\frac{d\sigma_{E}^{V_l}(\theta_{l},s,s')}{d\Omega}=
\frac{1}{2}\sigma^{l}_{o}(1-\eta^{2}_{l})^{-1}\{(1+ss')f_{1l}^{2}+$$
\begin{equation}
+\eta^{2}_{l}(1-ss')[f_{1l}^{2}+
4(m_{l}^{norm})^{2}(1-\eta^{-2}_{l})^{2}f_{2l}^{2}]
tg^{2}\frac{\theta_{l}}{2}\} F^{2}_{E}(q^{2}),
\label{19}
\end{equation}
where it has been accepted that
$$\sigma_{o}^{l}=
\frac{\alpha^{2}cos^{2}(\theta_{l}/2)}{4E^{2}_{l}(1-\eta^{2}_{l})
sin^{4}(\theta_{l}/2)}, \, \, \, \, \eta_{l}=
\frac{m_{l}^{norm}}{E_{l}},$$
$$E_{l}=\sqrt{p^{2}+(m_{l}^{norm})^{2}}, \, \, \, \,
F_{E}(q^{2})=ZF_{c}(q^{2}),$$
$$q^{2}=-4E_{l}^{2}(1-\eta_{l}^{2})sin^{2}\frac{\theta_{l}}{2}.$$
Here $\theta_{l}$ is the scattering angle, $E_{l}$ and $m_{l}$ are the
particle normal electric mass and energy, $F_{c}(q^{2})$ is the nucleus
charge $(F_{c}(0)=1)$ form factor. The index $V_{l}$ implies the absence
of the neutrino axial-vector $A_{l}$ currents.

The availability of terms $(1+ss')$ and $(1-ss')$ indicates to the existence of
vector Coulomb interactions of the left $(s=-1)$ and right $(s=+1)$ polarized
neutrinos, such as the conserving $(s'=s)$ and changing $(s'=-s)$ their
helicity. It is convenient therefore to reduce (\ref{19}) to the following:
\begin{equation}
d\sigma_{E}^{V_{l}}(\theta_{l},s)=
d\sigma_{E}^{V_{l}}(\theta_{l},f_{1l},s)+
d\sigma_{E}^{V_{l}}(\theta_{l},f_{2l},s),
\label{20}
\end{equation}
$$\frac{d\sigma_{E}^{V_{l}}(\theta_{l},f_{1l},s)}{d\Omega}=
\frac{d\sigma_{E}^{V_{l}}(\theta_{l},f_{1l},s'=s)}{d\Omega}+
\frac{d\sigma_{E}^{V_{l}}(\theta_{l},f_{1l},s'=-s)}{d\Omega}=$$
\begin{equation}
=\sigma^{l}_{o}(1-\eta^{2}_{l})^{-1}(1+\eta_{l}^{2}tg^{2}
\frac{\theta_{l}}{2})f_{1l}^{2}F_{E}^{2}(q^{2}),
\label{21}
\end{equation}
\begin{equation}
\frac{d\sigma_{E}^{V_{l}}(\theta_{l},f_{2l},s)}{d\Omega}=
\frac{d\sigma_{E}^{V_{l}}(\theta_{l},f_{2l},s'=-s)}{d\Omega}=
4(m_{l}^{norm})^{2}\sigma^{l}_{o}(1-\eta^{2}_{l})\eta^{-2}_{l}f_{2l}^{2}
F_{E}^{2}(q^{2})tg^{2}\frac{\theta_{l}}{2}.
\label{22}
\end{equation}

Making the averaging over $s$ and summing over $s',$ we can also present (\ref{19}) in the form
\begin{equation}
d\sigma_{E}^{V_{l}}(\theta_{l})=
d\sigma_{E}^{V_{l}}(\theta_{l},f_{1l})+
d\sigma_{E}^{V_{l}}(\theta_{l},f_{2l}),
\label{23}
\end{equation}
\begin{equation}
\frac{d\sigma_{E}^{V_{l}}(\theta_{l},f_{1l})}{d\Omega}=
\sigma^{l}_{o}(1-\eta^{2}_{l})^{-1}
(1+\eta^{2}_{l}tg^{2}\frac{\theta_{l}}{2})f_{1l}^{2}F_{E}^{2}(q^{2}),
\label{24}
\end{equation}
\begin{equation}
\frac{d\sigma_{E}^{V_{l}}(\theta_{l},f_{2l})}{d\Omega}=
4(m_{l}^{norm})^{2}\sigma^{l}_{o}(1-\eta^{2}_{l})\eta^{-2}_{l}f_{2l}^{2}
F_{E}^{2}(q^2)tg^{2}\frac{\theta_{l}}{2}.
\label{25}
\end{equation}

Thus, it follows that (\ref{19}) constitutes the two classes of vector cross sections, namely
\begin{equation}
d\sigma_{E}^{V_{l}}(\theta_{l},s)=
\{d\sigma_{E}^{V_{l}}(\theta_{l},f_{1l},s), \, \, \,\,
d\sigma_{E}^{V_{l}}(\theta_{l},f_{2l},s)\},
\label{26}
\end{equation}
\begin{equation}
d\sigma_{E}^{V_{l}}(\theta_{l})=
\{d\sigma_{E}^{V_{l}}(\theta_{l},f_{1l}), \, \, \, \,
d\sigma_{E}^{V_{l}}(\theta_{l},f_{2l})\}.
\label{27}
\end{equation}

Comparing (\ref{20}) and (\ref{23}), it is not difficult to see that
\begin{equation}
\frac{d\sigma_{E}^{V_{l}}(\theta_{l},s)}
{d\sigma_{E}^{V_{l}}(\theta_{l})}=1.
\label{28}
\end{equation}

Insertion of (\ref{20}) and (\ref{23}) in (\ref{28}) leads us to the conclusion that the 
interratios of possible pairs of elements from the sets (\ref{26}) and (\ref{27}) allow to 
establish the six different ratios. Among them one can meet the following:
\begin{equation}
\frac{d\sigma_{E}^{V_{l}}(\theta_{l},f_{1l},s)}
{d\sigma_{E}^{V_{l}}(\theta_{l},f_{2l})}, \, \, \, \,
\frac{d\sigma_{E}^{V_{l}}(\theta_{l},f_{2l},s)}
{d\sigma_{E}^{V_{l}}(\theta_{l},f_{1l})},
\label{29}
\end{equation}
\begin{equation}
\frac{d\sigma_{E}^{V_{l}}(\theta_{l},f_{2l},s)}
{d\sigma_{E}^{V_{l}}(\theta_{l},f_{1l},s)}, \, \, \, \,
\frac{d\sigma_{E}^{V_{l}}(\theta_{l},f_{2l})}
{d\sigma_{E}^{V_{l}}(\theta_{l},f_{1l})}.
\label{30}
\end{equation}

As well as the availability of the two remaining solutions, the equality to unity of each 
of these ratios is by no means excluded naturally.

Here it is relevant to remind once more about the presence in (\ref{19}) of the term $(1-ss')$ 
which assumed that the right-handed neutrinos similarly to the left-handed ones can essentially interact with the field of emission. As a consequence, the possibility of the existence in the 
field of a nucleus of interconversions $\nu_{L}\leftrightarrow \nu_{R}$ and
${\bar \nu}_{R}\leftrightarrow {\bar \nu}_{L}$ is nohow excluded. However, according to 
the standard model [16-18] based on an electroweak group $SU(2)_{L}$ $\otimes$ $U(1),$
the $\nu_{L}\nu_{R}$-interaction arising at the expense of fermion mass in the neutrino does not exist. Therefore, to establish the full spin structure of the above-mentioned system of equations 
and to use the neutrino helicity conservation or nonconservation in studying its properties one 
must elucidate the nature of the corresponding mass responsible for these transitions.

But at a given stage we can start from
\begin{equation}
\frac{d\sigma_{E}^{V_{l}}(\theta_{l},f_{2l},s)}
{d\sigma_{E}^{V_{l}}(\theta_{l},f_{1l},s)}=1, \, \, \, \,
\frac{d\sigma_{E}^{V_{l}}(\theta_{l},f_{2l})}
{d\sigma_{E}^{V_{l}}(\theta_{l},f_{1l})}=1.
\label{31}
\end{equation}

These equalities is a consequence of the unified regularity of Coulomb nature of leptons that 
the usual Dirac interaction comes forward in the system as the source of a non-zero interaction 
of Pauli. In other words, any of (\ref{31}) is valid only for those particles which possess simultaneously both charge and vector dipole moment.

At the explicit values of cross sections (\ref{21}), (\ref{22}), (\ref{24})
and (\ref{25}), they suggest the following connection of parameters
\begin{equation}
4(m_{l}^{norm})^{2}\frac{f_{2l}^{2}(0)}{f_{1l}^{2}(0)}=
\frac{\eta_{l}^{2}(1+\eta_{l}^{2}tg^{2}(\theta_{l}/2))}
{(1-\eta_{l}^{2})^{2}tg^{2}(\theta_{l}/2)}.
\label{32}
\end{equation}

Turning to (\ref{32}) and choosing the limit $E_{l}\gg m_{l},$ at which $\eta_{l}\rightarrow 0,$ 
and $q^{2}\rightarrow 0$ says $\theta_{l}\rightarrow 0$ we get, after the disclosure of 
uncertainties, such as
$$lim_{\eta_{l}\rightarrow 0, \theta_{l}\rightarrow 0}
\frac{\eta_{l}^{2}(1+\eta_{l}^{2}tg^{2}(\theta_{l}/2))}
{(1-\eta_{l}^{2})^{2}tg^{2}(\theta_{l}/2)}=1,$$
the corresponding dependence of form factors
\begin{equation}
2m_{l}^{norm}\frac{f_{2l}(0)}{f_{1l}(0)}=\pm 1.
\label{33}
\end{equation}

Thus, the absence of one of functions $f_{1l}(0)$ or $f_{2l}(0)$ implies that both do not 
exist at all.

\vspace{0.6cm}
\noindent
{\bf 3. United Dependence of Vector Form Factors of Neutrino and Electron}
\vspace{0.2cm}

We have seen that between the structural parts of Dirac and Pauli vector currents of leptons there exists an individual connection [5]. Such a regularity, however, encounters a problem which states that an equality to unity of each of ratios (\ref{29}) and (\ref{30}) is strictly nonverisimilar. Thus, an equation (\ref{33}) implied from (\ref{31}) would seem does not correspond to the reality at all. It can be easily convinced, however, that this is not so.

Here an important circumstance is the fact that regardless of sizes of the corresponding 
cross sections, any of ratios (\ref{29}) and (\ref{30}) for the lepton and its neutrino 
coincides. But all of them cannot be used for any physically definite purposes. The point 
is that the same fermion may not be simultaneously both an unpolarized and a longitudinal 
polarized particle. Furthermore, if massive neutrinos are of left-helicities, they suffer 
a scattering either with or without flip of the spin.

But we can apply to (\ref{30}), namely to the equalities
\begin{equation}
\frac{d\sigma_{E}^{V_{\nu_{l}}}(\theta_{\nu_{l}},f_{2\nu_{l}},s)}
{d\sigma_{E}^{V_{\nu_{l}}}(\theta_{\nu_{l}},f_{1\nu_{l}},s)}=
\frac{d\sigma_{E}^{V_{l}}(\theta_{l},f_{2l},s)}
{d\sigma_{E}^{V_{l}}(\theta_{l},f_{1l},s)},
\label{34}
\end{equation}
\begin{equation}
\frac{d\sigma_{E}^{V_{\nu_{l}}}(\theta_{\nu_{l}},f_{2\nu_{l}})}
{d\sigma_{E}^{V_{\nu_{l}}}(\theta_{\nu_{l}},f_{1\nu_{l}})}=
\frac{d\sigma_{E}^{V_{l}}(\theta_{l},f_{2l})}
{d\sigma_{E}^{V_{l}}(\theta_{l},f_{1l})}.
\label{35}
\end{equation}

These connections arise because of that $f_{1l}(0)$ and $f_{2l}(0)$ correspond
to the two form of the same regularity of the nature of lepton normal Coulomb
interaction. In other words, the absence of one of them would imply that both
leptons and their neutrinos do not possess neither electric mass, charge nor
vector dipole moment.

Together with (\ref{21}), (\ref{22}), (\ref{24}) and (\ref{25}) they replace (\ref{32}) to
\begin{equation}
(m_{\nu_{l}}^{norm})^{2}\frac{f_{2\nu_{l}}^{2}(0)}{f_{1\nu_{l}}^{2}(0)}
\frac{\eta_{l}^{2}(1+\eta_{l}^{2}tg^{2}(\theta_{l}/2))}
{(1-\eta_{l}^{2})^{2}tg^{2}(\theta_{l}/2)}=
(m_{l}^{norm})^{2}\frac{f_{2l}^{2}(0)}{f_{1l}^{2}(0)}
\frac{\eta_{\nu_{l}}^{2}(1+\eta_{\nu_{l}}^{2}tg^{2}(\theta_{\nu_{l}}/2))}
{(1-\eta_{\nu_{l}}^{2})^{2}tg^{2}(\theta_{\nu_{l}}/2)}.
\label{36}
\end{equation}

By following the same arguments that allow to establish an individual dependence (\ref{33}), 
one can found from (\ref{36}) that
\begin{equation}
m_{\nu_{l}}^{norm}\frac{f_{2\nu_{l}}(0)}{f_{1\nu_{l}}(0)}=
\pm m_{l}^{norm}\frac{f_{2l}(0)}{f_{1l}(0)}.
\label{37}
\end{equation}

Unification of (\ref{37}) with (\ref{8}) at $l=\nu_{l}$ leads us to (\ref{33}) once more, 
confirming that the existence of individual as well as of united connections of the structural components of vector cross sections in the lepton and its neutrino scattering is, by itself, 
not excluded.

From the point of view of the standard electroweak theory of particles [16-18], the anomalous 
part of an electronic neutrino vector dipole moment
\begin{equation}
A_{2\nu_{e}}(0)=\frac{A_{1\nu_{e}}(0)}{2m_{\nu_{e}}^{anom}}
\label{38}
\end{equation}
arises as a result of one-loop phenomenon [19].

Its value is, according to our description, connected with the neutrino anomalous electric 
mass $m_{\nu_{e}}^{anom}$ and equal to
\begin{equation}
A_{2\nu_{e}}(0)=\mu_{\nu_{e}}^{anom}=
\frac{3eG_{F}m_{\nu_{e}}^{anom}}{8\pi^{2}\sqrt{2}}, \, \, \, \,
e=|e_{e}^{norm}|.
\label{39}
\end{equation}

On the basis of (\ref{38}) and (\ref{39}) it is obtained that the neutrino anomalous electric 
charge has the size
\begin{equation}
e_{\nu_{e}}^{anom}=-A_{1\nu_{e}}(0)=
-\frac{3eG_{F}(m_{\nu_{e}}^{anom})^{2}}{4\pi^{2}\sqrt{2}}.
\label{40}
\end{equation}

Here it is relevant to note once more that the electric mass and charge of a neutrino correspond 
to the two form of the same regularity of its Coulomb nature. By this reason we conclude that
\begin{equation}
m_{\nu_{e}}^{E}=m_{\nu_{e}}^{norm}+m_{\nu_{e}}^{anom}+...,
\label{41}
\end{equation}
\begin{equation}
e_{\nu_{e}}^{E}=e_{\nu_{e}}^{norm}+e_{\nu_{e}}^{anom}+....
\label{42}
\end{equation}

The solutions (\ref{39})-(\ref{42}) together with equations (\ref{6}) and (\ref{7}) give the possibility to understand the fact [5] that the neutrino full electric charge and vector dipole moment in the static limit behave as
\begin{equation}
e_{\nu_{e}}=e_{\nu_{e}}^{full}=
-\frac{3eG_{F}(m_{\nu_{e}}^{E})^{2}}{4\pi^{2}\sqrt{2}},
\label{43}
\end{equation}
\begin{equation}
\mu_{\nu_{e}}=\mu_{\nu_{e}}^{full}=
\frac{3eG_{F}m_{\nu_{e}}^{E}}{8\pi^{2}\sqrt{2}}.
\label{44}
\end{equation}

At first sight, unlike $\mu_{\nu_{e}},$ the size of $e_{\nu_{e}}$ contains not only 
the contributions of the normal and anomalous masses, but also the contribution of their interference between themselves. On the other hand, as follows from considerations of symmetry, 
the numbers of terms in (\ref{43}) and (\ref{44}) must coincide. At the same time, the neutrino itself can have simultaneously both normal and anomalous Coulomb interactions. This becomes 
possible owing to the unified nature of their structure.

If we now take into account that any of currents $F_{1l}(q^{2})$ and $F_{2l}(q^{2})$ arises at 
the availability of the electric mass, no doubt that $m_{l}$ is strictly 
multicomponent vector 
size [20]. We cannot therefore exclude the possibility that
\begin{equation}
m_{l}^{2}=|\vec{m_{l}^{E}}|^{2}=
(m_{l}^{norm})^{2}+(m_{l}^{anom})^{2}+....
\label{45}
\end{equation}

Such a picture leading to the flip of a neutrino spin [13,21] predicts the absence of the 
mixed-interference contribution of Dirac and Pauli vector form factors in all processes 
with these currents.

\vspace{0.6cm}
\noindent
{\bf 4. Conclusion}
\vspace{0.2cm}

Our analysis shows that between the normal parts of the electric charge and vector dipole moment 
of any lepton and its neutrino there exists an individual [5] as well as the united dependence. These connections, however, encounter many problems which require the elucidation of the ideas 
of each of them.

According to one of the dynamical aspects of mass-charge duality [4], the force of gravity of 
the Newton $F_{N}$ between the two neutrinos may be expressed through the force of the Coulomb $F_{C}$ among these particles and vice versa. Uniting their expression with (\ref{43}) and 
taking into account that $F_{C}>F_{N},$ we get [5]
\begin{equation}
m_{\nu_{e}}^{E}> 1.53\cdot 10^{-3}\ {\rm eV},
\label{46}
\end{equation}
\begin{equation}
e_{\nu_{e}}^{E}> 1.46\cdot 10^{-30}\ {\rm e}.
\label{47}
\end{equation}

Basing on the comparatively new $\beta$-decay experiment, it was found [22,23] that $m_{\nu_{e}}<2.5\ {\rm eV}.$ Insertion of this value in (\ref{43}) gives 
$e_{\nu_{e}}< 3.92\cdot 10^{-24}\ {\rm e}.$ These sizes together with (\ref{46}) and 
(\ref{47}) predict the following restrictions on the neutrino electric mass and charge:
\begin{equation}
1.53\cdot 10^{-3}\ {\rm eV}< m_{\nu_{e}}^{E}< 2.5\ {\rm eV},
\label{48}
\end{equation}
\begin{equation}
1.46\cdot 10^{-30}\ {\rm e}< e_{\nu_{e}}^{E}< 3.92\cdot 10^{-24}\ {\rm e}.
\label{49}
\end{equation}

The presented here discrepancies are not casual. They reflect the highly characteristic 
features of the compound structure of mass and charge that there are many uncertainties 
in the behavior as well as in the size of the experimentally observed properties 
of the neutrino.

Returning to (\ref{29}) and (\ref{30}), we remark that their equality in the processes both 
with lepton and with its neutrino corresponds in nature to the definite type of lepton numbers. 
In other words, each of them is valid only for particles of the same flavors.

According to this thought, the structural connection (\ref{37}) reflects the coexistence of 
leptons and their neutrinos. As a consequence, any of possible types $(l=e,$ $\mu,$ $\tau, ...)$ 
of charged leptons testifies in favor of the existence of a kind of Dirac 
$(\nu_{l}=\nu_{e},$ $\nu_{\mu},$ $\nu_{\tau}, ...)$ neutrino. They constitute therefore 
the naturally united families of leptons.

Such a regularity, however, takes place in force of the unified nature of each pair of fermions 
that the absence of any of the lepton and its neutrino, as stated in (\ref{37}), would imply that neither particle exists.

At this situation, the earlier measured properties of particles $e^{-}_{R}(e^{+}_{L}),$
$\mu^{-}_{R}(\mu^{+}_{L})$ and $\tau^{-}_{R}(\tau^{+}_{L})$ may serve as a certain indication
to the existence simultaneously of each of the right (left)-handed lepton 
(antilepton) and its neutrino (antineutrino) but not of one of them.

In a given case, from (\ref{37}) we are led to a unified principle that leptons and their 
neutrinos constitute the naturally united families both of the left-handed $SU(2)_{L}$-doublets 
and of the right-handed $SU(2)_{R}$-doublets:
\begin{equation}
\pmatrix{\nu_{e}\cr e^{-}}_{L},
(\nu_{e}, \, \, \, \, e^{-})_{R}, \, \, \, \,
\pmatrix{\nu_{\mu}\cr \mu^{-}}_{L},
(\nu_{\mu}, \, \, \, \, \mu^{-})_{R}, \, \, \, \,
\pmatrix{\nu_{\tau}\cr \tau^{-}}_{L},
(\nu_{\tau}, \, \, \, \, \tau^{-})_{R}, ...,
\label{50}
\end{equation}
\begin{equation}
\pmatrix{{\bar \nu_{e}}\cr e^{+}}_{R},
({\bar \nu_{e}}, \, \, \, \, e^{+})_{L}, \, \, \, \,
\pmatrix{{\bar \nu_{\mu}}\cr \mu^{+}}_{R},
({\bar \nu_{\mu}}, \, \, \, \, \mu^{+})_{L}, \, \, \, \,
\pmatrix{{\bar \nu_{\tau}}\cr \tau^{+}}_{R},
({\bar \nu_{\tau}}, \, \, \, \, \tau^{+})_{L}, ....
\label{51}
\end{equation}

This in turn predicts the availability in nature of a gauge group
\begin{equation}
SU(2)_{L}\otimes SU(2)_{R}\otimes U(1),
\label{52}
\end{equation}
describing the fundamentally important consequences of the micro-world symmetry. 
One of them is that the numbers of components in all types of massive particles 
must coincide [24].

To express the idea more clearly one must refer to the structural dependence (\ref{37}) 
which states that
\begin{equation}
m_{\nu_{l}}^{norm}f_{1l}(0)f_{2\nu_{l}}(0)-
m_{l}^{norm}f_{1\nu_{l}}(0)f_{2l}(0)=0.
\label{53}
\end{equation}

There exists, however, the possibility [21] that charge $f_{1l}(0),$ as follows from (\ref{21}), leads to the process going either with or without change of incoming particle helicities. Unlike this, the vector dipole moment $f_{2l}(0)$ is responsible only for the interconversion of longitudinal leptons of the different components. Therefore, if it turns out that the same 
fermion may not be simultaneously both a left-handed and a right-handed particle, each 
interference term in (\ref{53}) indicates to the existence of a kind of unified system 
of the two types of leptons of the same families of $SU(2)_{L}$- or $SU(2)_{R}$-doublets. 
One can call their the left (right) dileptons. They behave as
\begin{equation}
(l^{-}_{L}, {\bar \nu_{lR}}), \, \, \, \, (l^{-}_{R}, {\bar \nu_{lL}}),
\label{54}
\end{equation}
\begin{equation}
(l^{+}_{R}, \nu_{lL}), \, \, \, \, (l^{+}_{L}, \nu_{lR}).
\label{55}
\end{equation}

One of highly characteristic features of these types of systems is the existence in them of 
the unified force which unites the structural particles of each flavor symmetrical pair.

Thus, any of equations (\ref{37}) and (\ref{52}) describes a connected system of the two united systems of the two types of the left (right)-handed Dirac fermions of the most diverse vector currents. It of course can be called the left (right) paradilepton.

One can define its structure in the form of one of systems
\begin{equation}
\{(l^{-}_{L}, {\bar \nu_{lR}}),
(l^{+}_{R}, \nu_{lL})\}, \, \, \, \,
\{(l^{-}_{R}, {\bar \nu_{lL}}), (l^{+}_{L}, \nu_{lR})\}.
\label{56}
\end{equation}

Such paradileptons as well as the united dileptons (\ref{54}) and (\ref{55}) appear as a 
consequence of conservation of lepton flavors, for example, at the elastic scattering on a 
spinless nucleus [25], if an incoming particle flux includes both leptons and their neutrinos.

Insofar as a question about the appearance of united dileptons
\begin{equation}
(e^{-}_{L}, {\bar \nu_{eR}}), \, \, \, \, (e^{-}_{R}, {\bar \nu_{eL}}),
\label{57}
\end{equation}
\begin{equation}
(e^{+}_{R}, \nu_{eL}), \, \, \, \, (e^{+}_{L}, \nu_{eR})
\label{58}
\end{equation}
and paradileptons
\begin{equation}
\{(e^{-}_{L}, {\bar \nu_{eR}}), (e^{+}_{R}, \nu_{eL})\}, \, \, \, \,
\{(e^{-}_{R}, {\bar \nu_{eL}}), (e^{+}_{L}, \nu_{eR})\}
\label{59}
\end{equation}
at the $\beta$-decay is concerned, we must have in view of that
\begin{equation}
m_{\nu_{e}}^{norm}f_{1e}(0)f_{2\nu_{e}}(0)-
m_{e}^{norm}f_{1\nu_{e}}(0)f_{2e}(0)=0,
\label{60}
\end{equation}
owing to which, the electron number in each of processes (\ref{16}) and (\ref{17}) is 
strictly conserved. In this appears a great responsibility of a formation both of dileptons 
and of paradileptons for the legality of conservation of lepton flavors as well as of full 
lepton number.

We recognize that at the conservation of gauge invariance, the term $f_{1\nu_{l}}(0)$ in 
(\ref{18}) must be equal to zero. Thus, it would seem can be drawn an implication about 
the absence of the neutrino electric mass.

In conformity with the ideas of mass-charge duality [4], this would take place only in the case where a Dirac neutrino is absent. Nevertheless, if we suppose that $m_{\nu_{l}}^{norm}=0$ then (\ref{37}) would lead us to the conclusion about Majorana nature of any lepton mass. There are, however, many weighty arguments [26-29] in favor of self masses not only of leptons, but also 
of their neutrinos. Therefore, to (\ref{37}) one must apply as to one of further confirmations 
of a new structure of electromagnetic gauge invariance [14] depending on a particle mass and
testifying about that mirror symmetry is basically violated at the expense of mass [24]. At 
the availability of exactly the same mass, each of neutrinos similarly to a kind of charged 
lepton must possess a non-zero electric charge.

Because of this, the cross section (\ref{19}) contains as well as the terms 
$f_{1l}^{2}(0)$ describing the charge contributions. Taking $\sigma_{\mu\lambda}=
i[\gamma_{\mu},\gamma_{\lambda}]/2$ and demanding that the Hermiticity of the neutrino-photon 
vertex is absent even at $q^{2}<0,$ we would get the other cross section [13] for the elastic scattering. It should include also the interference term $f_{1l}(0)f_{2l}(0)$ which would characterize the existence in nature of each of paraparticles
\begin{equation}
(l_{L}^{-}, l_{R}^{+}), \, \, \, \, (l_{R}^{-}, l_{L}^{+}).
\label{61}
\end{equation}

It is clear, however, that the Hermiticity of leptonic current is not excluded. Insofar as 
the individual dileptons (\ref{60}) are concerned, their scattering on a spinless nucleus can 
be explained by a self interference contribution to the process cross section (\ref{19}) of any 
of the interaction Dirac $f_{1l}(0)$ and Pauli $f_{2l}(0)$ parts. In other words, these phenomena
originate in the flavor symmetrical field in which nobody is in force to observe simultaneously 
the same lepton regardless of a kind of antiparticle.

The above reasonings say about that unlike the standard electroweak model, the discussed theory 
of force unification based on a gauge group (\ref{52}) must be left-right antisymmetric theory including a unified theoretical description of all types of left- and right-handed fermions. Thereby, it predicts the availability in nature of an unbroken flavor symmetry. 

According to the charge conservation law, with its violation would originate an instantaneous reestablishment of the structure of the Coulomb field [6]. But in nature there is no such 
an order [30-32] in force of symmetry laws.

Conservation of charge comes forward in the system as a consequence of one of the most diverse symmetries [33] of gauge fields before and after the emission. Therefore, the legality of charge conservation at the neutrino interaction may be established only in the case where appear the dileptons and paradileptons. From this point of view, the earlier discovered in the experiences [34-36] processes, for example, (\ref{16}), (\ref{17}) and
\begin{equation}
\gamma e^{-}\rightarrow e^{-}\nu_{e}{\bar \nu_{e}}, \, \, \, \,
e^{-}e^{+}\rightarrow \nu_{e}{\bar \nu_{e}},
\label{62}
\end{equation}
\begin{equation}
\nu_{e}e^{-}\rightarrow \nu_{e}e^{-}, \, \, \, \,
{\bar \nu_{e}}e^{-}\rightarrow {\bar \nu_{e}}e^{-},
\label{63}
\end{equation}
conserving charge and lepton number may serve as the first confirmation
of the existence of each of dileptons and paradileptons.

Another consequence of the united connection (\ref{37}) is an interratio of masses of leptons 
of the same family:
\begin{equation}
\frac{m_{\nu_{l}}^{norm}}{m_{l}^{norm}}=
\frac{f_{1\nu_{l}}(0)}{f_{1l}(0)}
\frac{f_{2l}(0)}{f_{2\nu_{l}}(0)}.
\label{64}
\end{equation}

This together with the ideas of full lepton number conservation law gives the possibility 
to directly look on the nature of the neutrino mass
\begin{equation}
\frac{m_{\nu_{e}}^{norm}}{m_{e}^{norm}}=
\frac{m_{\nu_{\mu}}^{norm}}{m_{\mu}^{norm}}=
\frac{m_{\nu_{\tau}}^{norm}}{m_{\tau}^{norm}}
\label{65}
\end{equation}
and that, consequently, the suggested theory which unites the leptons and their neutrinos 
predicts the existence of one more highly important connection:
\begin{equation}
m_{\nu_{e}}^{norm}:m_{\nu_{\mu}}^{norm}:m_{\nu_{\tau}}^{norm}=
m_{e}^{norm}:m_{\mu}^{norm}:m_{\tau}^{norm}.
\label{66}
\end{equation}

In the same way one can see from (\ref{37}) that
\begin{equation}
e_{\nu_{e}}^{norm}:e_{\nu_{\mu}}^{norm}:e_{\nu_{\tau}}^{norm}=
e_{e}^{norm}:e_{\mu}^{norm}:e_{\tau}^{norm},
\label{67}
\end{equation}
\begin{equation}
\mu_{\nu_{e}}^{norm}:\mu_{\nu_{\mu}}^{norm}:\mu_{\nu_{\tau}}^{norm}=
\mu_{e}^{norm}:\mu_{\mu}^{norm}:\mu_{\tau}^{norm}.
\label{68}
\end{equation}

Therefore, we conclude that
\begin{equation}
m_{\nu_{e}}:m_{\nu_{\mu}}:m_{\nu_{\tau}}=
m_{e}:m_{\mu}:m_{\tau},
\label{69}
\end{equation}
\begin{equation}
e_{\nu_{e}}:e_{\nu_{\mu}}:e_{\nu_{\tau}}=
e_{e}:e_{\mu}:e_{\tau},
\label{70}
\end{equation}
\begin{equation}
\mu_{\nu_{e}}:\mu_{\nu_{\mu}}:\mu_{\nu_{\tau}}=
\mu_{e}:\mu_{\mu}:\mu_{\tau}.
\label{71}
\end{equation}

Thus, unlike the earlier presentations on the structure of leptonic families [37,38], the 
suggested theory of unification of leptons and their neutrinos leads us to a correspondence principle that the mass, charge and vector dipole moment of the neutrino are proportional, respectively, to the mass, charge and vector dipole moment of a lepton of the same family.

For completeness, we include in the discussion an experimental possibility to follow in the 
nucleus field the Coulomb behavior of massive Dirac neutrinos. In the case of spinless nuclei, 
on the elastic scattering influence only the properties of an incoming particle itself. With 
this was connected the fact [39] that the scattering angle $\theta_{l}$ may be expressed as a function of the neutrino charge $f_{1l}(0)$ and vector dipole $f_{2l}(0)$ form factors:
\begin{equation}
\theta_{l}=\pm 2Arctg \{4E_{l}^{2}(1-\eta_{l}^{2})^{2}
\frac{f_{2l}^{2}(0)}{f_{1l}^{2}(0)}-\eta_{l}^{2}\}^{-1/2}.
\label{72}
\end{equation}

It states that high energy leptons $(E_{l}\gg m_{l}),$ for which $\eta_{l}\rightarrow 0,$ 
suffer the scattering almost forward $(\theta_{l}\rightarrow 0),$ namely
\begin{equation}
\theta_{l}=\pm 2Arctg\frac{f_{1l}(0)}{2E_{l}f_{2l}(0)}.
\label{73}
\end{equation}

In these conditions, $q^{2}\rightarrow 0,$ and the cross section of unpolarized fermion 
scattering (\ref{23}) with the account of $f_{1l}^{2}(0)/E_{l}^{2}\rightarrow 0$ has 
the following structure:
\begin{equation}
\frac{d\sigma_{em}^{V_{l}}(\theta_{l})}{d\Omega}=
\frac{Z^{2}\alpha^{2}}{sin^{2}(\theta_{l}/2)}f_{2l}^{2}(0).
\label{74}
\end{equation}

It is clear that measurement of $\theta_{l}$ for any two values of the fast lepton energies 
allows to define the new laboratory sizes of the neutrino charge and vector dipole moment [39]. Thereby, one can estimate the cross section (\ref{74}) as well as the fine-structure constant
$\alpha,$ whose value has not yet been obtained in the nucleus Coulomb field with neutrinos.

At such studies, it should be chosen only a nucleus with zero spin and isospin (for example, 
from $^{4}He,$ $^{12}C,$ $^{40}Ca,$ ...), so that the target nucleus isotopic structure strongly changes the neutrino properties [40]. Of course, similar measurements require the high sensitivity of devices confirming that a nucleus comes forward in the system as an innate supersensitive detector for observation of fundamental symmetry laws.

\vspace{0.6cm}
\noindent
{\bf References}
\begin{enumerate}
\item
M.N. Rosenbluth, Phys. Rev. {\bf 79} (1950) 615.
\item
R.S. Sharafiddinov, Spacetime Subst. {\bf 3} (2002) 132; physics/0305014.
\item
R.S. Sharafiddinov, Spacetime Subst. {\bf 3} (2002) 86; physics/0305009.
\item
R.S. Sharafiddinov, Spacetime Subst. {\bf 3} (2002) 47; physics/0305008.
\item
R.S. Sharafiddinov, Spacetime Subst. {\bf 1} (2000) 176; hep-ph/0305009.
\item
Ya.B. Zel'dovich and M.Yu. Khlopov, Uspehi Fiz. Nauk. {\bf 135} (1981) 45.
\item
Ya.B. Zel'dovich, Dokl. Akad. Nauk SSSR. {\bf 86} (1952) 505.
\item
E.J. Konopinsky and H. Mahmoud, Phys. Rev. {\bf 92} (1953) 1045.
\item
A. Van Der Schaaf, in {\it Proceedings of the Summer School on Particle
Physics, Zuoz, August 18-24, 2002} (Z\"urich, Switzerland, 2003), p. 269.
\item
J. Kubo et al., Prog. Theor. Phys. {\bf 109} (2003) 795; hep-ph/0302196.
\item
T.P. Cheng and Lee Ling-Fond, Phys. Rev. Lett. {\bf 38}
(1977) 381.
\item
B.W. Lee et al., Phys. Rev. Lett. {\bf 38} (1977) 937.
\item
R.S. Sharafiddinov, Dokl. Akad. Nauk Ruz. Ser. Math. Tehn. Estest. {\bf 7} (1998) 25; 
hep-ph/0307083.
\item
R.S. Sharafiddinov, Spacetime Subst. {\bf 5} (2004) 83;
hep-ph/0306255.
\item
T.W. Donnelly and R.D. Peccei, Phys. Rep. {\bf 50} (1979) 3.
\item
S.L. Glashow, Nucl. Phys. {\bf 22} (1961) 579. 
\item
A. Salam and J.C. Ward, Phys. Lett. {\bf 13} (1964) 168. 
\item
S. Weinberg, Phys. Rev. Lett. {\bf 19} (1967) 1264.
\item
K. Fujikawa and R.E. Shrock, Phys. Rev. Lett. {\bf 45} (1980) 963.
\item
R.S. Sharafiddinov, Spacetime Subst. {\bf 4} (2003) 79; hep-ph/0401230.
\item
R.S. Sharafiddinov, Spacetime Subst. {\bf 3} (2002) 134; physics/0305015.
\item
C.V. Weinheimer et al., Phys. Lett. {\bf B 460} (1999) 219.
\item
V.M. Lobashev et al., Phys. Lett. {\bf B 460} (1999) 227.
\item
R.S. Sharafiddinov, Phys. Essays {\bf 19} (2006) 58; hep-ph/0407262.
\item
R.S. Sharafiddinov, in {\it Proceedings of the 2rd Eurasian Conference on Nuclear Science 
and Its Application, Almaty, September 16-19, 2002} (Almaty, Kazakhstan, 2002), 
Abstracts, p. 146.
\item
P.Q. Hung, Phys. Rev. {\bf D 67} (2003) 095011; hep-ph/0210131.
\item
M. Jezabek and P. Urban, Phys. Lett. {\bf B 541} (2002) 142; hep-ph/0206080.
\item
P.H. Frampton and P. Vogel, Phys. Rep. {\bf 82} (1982) 339.
\item
F. Boehm and P. Vogel, Ann. Rev. Nucl. Part. Sci. {\bf 34} (1984) 125.
\item
L.B. Okun and Ya.B. Zel'dovich, Phys. Lett. {\bf B 78}
(1978) 597.
\item
B.A. Kouz'min and M.J. Shaposhnikov, Pis'ma v Zh. Eksp. Teor. Fiz. {\bf 27} (1978) 665.
\item
M.B. Voloshin and L.B. Okun, Pis'ma v Zh. Eksp. Teor. Fiz. {\bf 28} (1978) 156.
\item
R.J. Ellis, in {\it Proceedings of the Summer School on Particle Physics, Zuoz, 
August 18-24, 2002} (Z\"urich, Switzerland, 2003), p. 1; hep-ph/0211168.
\item
J. Bernstein, M. Ruderman and G. Feinberg, Phys. Rev. {\bf 132} (1963) 1227.
\item
H.S. Gurr, F. Reines and H.W. Sobel, Phys. Rev. Lett. {\bf 28} (1972) 1406.
\item
S. Davidson, B. Campbell and K.D. Bailey, Phys. Rev. {\bf D 43} (1991) 2314.
\item
M. Gell-Mann et. al., in Supergravity, Amsterdam, North-Holland, 1979.
\item
T. Yanagida, Workshop on Unified Theory and Baryon Number of Universe, unpublished, 1979.
\item
R.B. Begzhanov and R.S. Sharafiddinov, in {\it Proceedings of the International Conference 
on Nuclear Physics, St-Petersburg, June 14-17, 2000} (St-Petersburg, 2000), p. 118.
\item
R.B. Begzhanov, R.S. Sharafiddinov, in {\it Proceedings of the International Conference 
on Nuclear Physics, Moscow, June 16-19, 1998} (St-Petersburg, 1998) Abstracts, p. 354.
\end{enumerate}
\end{document}